\documentclass[a4paper,fleqn,usenatbib]{mnras}
\usepackage[T1]{fontenc}
\usepackage{ae,aecompl}
\usepackage{graphicx}
\usepackage{amsmath}
\usepackage{amssymb}
\usepackage{ulem}

\title[Simulations of a possible ring around G\,196-3\,B]
{Spectral energy distribution simulations of a possible ring  structure around the 
young, red brown dwarf G\,196-3\,B}
\author[O. V. Zakhozhay et al.]{Olga V. Zakhozhay$^1$\thanks{E-mail:zakhozhay.olga@gmail.com},
 Mar\'{\i}a Rosa Zapatero Osorio$^2$, V\'{\i}ctor J. S. B\'{e}jar$^{3,4}$, 
\newauthor and Yann Boehler$^5$ \\
$^{1}$ Main Astronomical Observatory, National Academy of Sciences of Ukraine, Kyiv, 03680, Ukraine \\
$^{2}$ Centro de Astrobiolog\'{\i}a (CSIC-INTA), Crta. Ajalvir km 4, E-28850 Torrej\'{o}n de Ardoz, Madrid, Spain \\
$^{3}$ Instituto de Astrof\'{\i}sica de Canarias (IAC), C. V\'{\i}a L\'{a}ctea s/n, E-38205 La Laguna, Tenerife, Spain \\
$^{4}$ Universidad de La Laguna, Dpto. Astrof\'{\i}sica, E38206 La Laguna, Tenerife, Spain \\
$^{5}$ Rice University, 6100 Main Street, Houston, TX, 77005, USA\\
}

\date{Accepted XXX. Received YYY; in original form ZZZ}

\pubyear{2016}

\begin{document}
\label{firstpage}
\pagerange{\pageref{firstpage}--\pageref{lastpage}}
\maketitle

\begin{abstract}
The origin of the very red optical and infrared colours of intermediate-age ($\sim$10--500~Myr) L-type dwarfs remains unknown. It has been suggested that low-gravity atmospheres containing large amounts of dust may account for the observed reddish nature. We explored an alternative scenario by simulating protoplanetary and debris discs around G\,196-3\,B, which is  an L3 young brown dwarf with a mass of $\sim 15$~$M_{\rm Jup}$ and an age in the interval 20--300~Myr. The best-fit solution to G\,196-3\,B's photometric spectral energy distribution from optical wavelengths through 24~$\mu$m corresponds to the combination of an unreddened L3 atmosphere ($T_{\rm eff} \approx 1870$~K) and a warm ($\approx$~1280~K), narrow ($\approx$~{0.07--0.11}~R$_{\odot}$) debris disc located at very close distances ($\approx$~0.12--0.20~R$_{\odot}$) from the central brown dwarf. This putative,  optically thick, dusty belt, whose presence is compatible with the relatively young system age, would have a mass $\ge 7\times 10^{-10}$~M$_{\oplus}$ comprised of sub-micron/micron characteristic dusty particles with temperatures close to the sublimation threshold of silicates. Considering the derived global properties of the belt and the disc-to-brown dwarf mass ratio, the dusty ring around G\,196-3\,B may resemble the rings of Neptune and Jupiter, except for its high temperature and thick vertical height ($\approx 6 \times 10^3$~km). Our inferred debris disc model is able to reproduce G\,196-3\,B's spectral energy distribution to a satisfactory level of achievement.
\end{abstract}


\section{Introduction}
\label{introduction}

To date, more than 1000 L-, T-, and  the recently discovered Y-type dwarfs have been {identified over the past 15 years (e.g. \citealt{Kirk99, Kirk00, Kirk11, Kirk13, martin99, Burgasser99, cushing11, tinney12, beichman13, Pinfield14, masters12}). About 10~per~cent of the field L dwarfs show spectroscopic signatures of low-to-intermediate} gravity and youth  (\citealt{mcgovern04,Kirk06, Cruz09}), usually accompanied by near- and mid-infrared colours redder than those of normal field sources of similar spectral classification (e.g. \citealt{Looper08,ZO10, Gizis12,Fah13,Liu13,Sch14,gauza15,gizis15}). Although novel synthetic spectra tend to explain these infrared flux excesses by the combining effects of low pressure conditions and thick dust clouds of the atmospheres (e.g. \citealt{barman11,marley12}), it is a fact that not all young and low gravity L dwarfs show markedly red indices, in particular at the very young ages ($\le$10 Myr, e.g. \citealt{martin01a,Lodieu08,pena15}), and not all very red L dwarfs display the full list of spectroscopic signposts indicative of low gravity (e.g. \citealt{marocco14,kellogg15}). In \cite{osorio14}, the authors derived ages in the interval $\approx$~10--500~Myr for a sample of field, highly reddened L dwarfs with clear spectroscopic evidence for low-to-intermediate gravity atmospheres. According to evolutionary models (e.g. \citealt{Chabrier00}), these dwarfs likely have substellar masses below the hydrogen burning-mass limit at 0.072~M$_{\odot}$ (for solar metallicity). 

Free-floating brown dwarfs in young star forming regions show the typical signatures associated with the presence of protoplanetary discs of gas and dust (strong and broad H$\alpha$ and other emission lines, marked infrared flux excesses), from which some objects are actually accreting material (e.g. \citealt{jayawardhana01,muzerolle03a}; see review by \citealt{luhman12}). Also, very young,  isolated planetary-mass objects below the deuterium burning mass limit at 0.012~M$_\odot$ ($\approx 13$~$M_{\rm Jup}$) can host primordial discs (e.g. \citealt{luhman05,luhman08,osorio07}). The  modelling of the observed spectral energy distributions suggests that the properties and grain growth of the circum(sub)stellar discs are quite similar to those related to T\,Tauri stars, scaled down to the smaller masses of the central objects (e.g. \citealt{Natta2001,Apai2008,Ricci2012,Ricci2013,Ricci2014}). \cite{luhman12b} concluded that primordial discs lifetimes are larger at lower stellar masses, and that a significant fraction of gas discs of low-mass objects survive for at least $\approx$~10--20~Myr (see also \citealt{Riaz09} and \citealt{ribas14}). After the gas is cleared out, dust accretes into planets and moons and/or remains as debris material circling the central object. According to the {\sl Herschel} observations conducted by \citet{riaz14}, the discs surrounding very young free-floating brown dwarfs appear to have transitioned from primordial to the debris phase by the age of $\sim$40--50~Myr. Therefore, debris discs are thought to occur around brown dwarfs and planetary-mass objects, although none has been unambiguously confirmed so far, partly because the dwarfs' low luminosity makes the thermal emission and scattered light from their discs more difficult to detect. Also, brown dwarf and planetary-mass companions to stars are capable of sustaining their own discs (see \citealt[and references therein]{Wu15}). \cite{mamajek12} and \cite{werkhoven14} suggested that the 16-Myr old star 1SWASP\,J140747.93$-$394542.6 has an unresolved substellar companion with a large ring system comprised of dusty particles, which may be responsible for the complex series of eclipses seen in the stellar light curve. On the contrary, the study of the {\sl Kepler} light curves of 21 hot-Jupiter planets, most likely with ages of Gyr, revealed no evidence for ringed exoplanets \citep{Heising15}.

Here, we provide an alternative interpretation for the very red near- and mid-infrared colours of the young, early-L dwarf G\,196-3\,B \citep{Rebolo98}, which is conceptually different to the proposed low gravity, enshrouded atmosphere. G\,196-3\,B is separated by $\approx$~390~au from the primary, active M2.5 star G\,196-3\,A. As discussed by \cite{Rebolo98} and \cite{ZO10,osorio14}, the system has an age in the interval 20--300~Myr, with a probable value around 50--90~Myr. The substellar companion, which has a mass of $15^{+30}_{-4}$~$M_{\rm Jup}$ near the deuterium-burning mass limit, shows redder colours at all wavelengths from 1.6 up to 24 $\mu$m than expected for its spectral type determined at L1--L3 from optical and near-infrared spectra \citep{martin99,kirk01,Cruz09,allers13}. The working hypothesis is that, given the relatively young age, G\,196--3\,B's flux excesses may mostly originate in a surrounding disc. We fit the observed photometric energy distribution from the optical through 24~$\mu$m using a debris disc model.  Our objectives are to explore whether physically possible fits can be obtained and to discuss a likely picture that accounts for the reddish properties of G\,196-3\,B. With plain assumptions on disc geometry and disc plane orientation, we found that a ring (debris) model reproduces many of the observed infrared fluxes. The paper is organised as follows: In Section~\ref{model}, we describe the modelling procedure for the disc. Results and their implications for the properties of G\,196-3\,B are presented in Section~\ref{results}. Finally, our conclusions and final remarks appear in Section~\ref{conclusions}.

\section{Disc modelling approach}
\label{model}
For the following analysis, we consider the photometric spectral energy distribution of G\,196-3\,B shown in \citet{ZO10} to be composed of a photospheric emission typical of an L2--L3 dwarf with no flux excesses, and an additional warm gas/dust component that may account for the observed red colours at long wavelengths.

One input parameter for any disc model is the luminosity of the central object, $L_{\ast}$, that heats and illuminates the disc. We calculated $L_{\ast}$ from the average spectral energy distribution (SED) of field L2--L3 sources (with no obvious colour anomalies) normalized to the $J$-band luminosity of G\,196-3\,B, which was presented in fig$.$~3 of \citet{ZO10}. This SED covers from the $R$-band up to 24~$\mu$m using the same filter passbands and central wavelengths as for G\,196-3\,B. By integrating the following expression over the extended wavelength range 0.3--100~$\mu$m (Simpson's rule):
\begin{equation}\label{form1}
 L_{\ast} = 4 \pi d^2 \int_{0}^{+\infty}f_{\nu,\ast} ~{\rm d}\nu,
  \end{equation}
where $d$ = 24.4$^{+2.7}_{-2.2}$~pc (parallactic distance to G\,196-3\,B, \citealt{osorio14}), $\nu$ is frequency, and $f_{\nu,\ast}$ is the observed photospheric flux density at each frequency, we obtained $L_{\ast} = 1.16^{+0.34}_{-0.24}\times10^{-4}$~L$_{\odot}$, where L$_{\odot}$ stands for the solar luminosity. At this point, we did not make any correction for interstellar extinction since the system G\,196-3 is located at a near distance and does not belong to any obscured star-forming region. The luminosity uncertainty is dominated by the distance error. Within the quoted error bar, this absolute luminosity coincides with the bolometric magnitudes reported for L3-type sources by \citet{dahn02} and with the luminosities of L3 dwarfs given by \citet{vrba04} and \citet{dieterich14}. However, it contrasts with the higher luminosity of G\,196-3\,B, $L = 1.86^{+0.70}_{-0.51}\times 10^{-4}$~L$_{\odot}$ (see table~6 of \citealt{osorio14}). According to our initial hypothesis, there might be an extra source of luminosity in the form of a disc.

In a first attempt, as the disc is not spatially resolved by our observations, we will consider it face-on. The SED of the disc is generated from the thermal reprocessing of the light of the central source by circum(sub)stellar material according to the following expression:
\begin{equation}\label{form2}
f_{\nu,disc}=d^{-2}\int_{R_{in}}^{R_{out}}B_{\nu}(T_r)~Q_{\nu}~2\pi r~{\rm d}r,
\end{equation}
\begin{equation}\label{form3}
Q_{\nu} = 1 - {\rm e}^{-\tau} ; ~~~ \tau = \varSigma_r~\kappa_{\nu},
\end{equation}
where $f_{\nu,disc}$ is the disc flux density for the frequency $\nu$, $r$ is the radial distance from the central object, $B_{\nu}(T_r)$ is the Planck function for a radius-dependent temperature, $T_r$, of the disc, and $R_{in}$ and $R_{out}$ stand for the inner and outer radii of the disc, respectively. $Q_{\nu}$ represents the radiative efficiency of the disc material, $\tau$ stands for the optical depth of the disc, which is the product of the wavelength-dependent disc opacity, $\kappa_{\nu}$, and the radial surface density distribution of the disc material, $\varSigma_r$. 

In our study, and for simplicity, we made the following assumptions: {\sl (i)} That there is no gas in the disc, which is consistent with the most likely age of G\,196-3\,B (50--90~Myr). This is we expect that the original amount of gas is dissipated and only solid remnants remain forming a debris disc. {\sl (ii)} That these solids can be parametrised by a single characteristic spherical grain size, $a$. And {\sl (iii)}, that the Mie mathematical-physical theory is valid to calculate the wavelength-dependent opacity properties of compact spherical particles (e.g., astronomical silicates with density $=~2.5$ g\,cm$^{-3}$) for the wavelength coverage of our analysis.

\par The disc surface density is described by the equation:
\begin{equation}\label{form9}
\varSigma_r=\varSigma_{in}\left(\frac{r}{R_{in}}\right)^{p}, 
\end{equation}
where $\varSigma_{in}$ is a surface density at $R_{in}$. Then the total mass of the disc is related to the disc size as follows:
\begin{equation}\label{md}
M_d = \int_{R_{in}}^{R_{out}}2 \pi r~\varSigma_r~{\rm d}r.
\end{equation}
Conversely, one can express the disc radial surface density as a function of the disc mass:
\begin{equation}\label{Sigma_r}
\varSigma_r = \frac {M_d~r^p~(2+p)}{2 \pi~(R_{out}^{2+p} - R_{in}^{2+p})},
\end{equation}
where $p$ is the index of the radial surface profile described in equation~\ref{form9}, and has a value of $p = -1.5$ for collisionally dominated discs. This is also the value accepted for describing debris discs in the solar system \citep{Carpenter09}.

\par We imposed that the grains at the inner edge of the disc must be in radiative equilibrium with the central substellar object, and that the disc is spatially extended.  Beyond the inner rim, the temperature of the disc is supposed to be a function of the distance to the central source as given by the following expression:

\begin{equation}\label{form10}
T_r=T_{in}\left(\frac{r}{R_{in}}\right)^{q}, 
\end{equation}
where $T_{in}$ corresponds to the radiative equilibrium temperature of the disc at $R_{in}$ and the index $q$ is taken to be a free parameter. We computed $T_{in}$ using the prescription of \cite{Backman93}, where the grain temperature is calculated from equilibrium between absorbed and emitted energy and depends on the ratio between the absorption and emission coefficients. Based on the L-dwarf and disc spectra presented in figure~6 of \citealt{ZO10}, we determined grains of $a\leq0.2$~$\mu$m as small grains, $a\geq2$~$\mu$m as large grains and in between as intermediate grains and computed $R_{in}$ with equations~6, 3 and 5 of \cite{Backman93}, respectively. We adopted a  characteristic wavelength $\lambda_0 = \pi a$ in equation 5 of \cite{Backman93}, which corresponds to moderately absorbing dielectrics.

\begin{table*}
\begin{minipage}{148 mm}
\begin{center}
\caption{Best-fit face-on disc parameters$^a$.}
\label{table1}
\renewcommand{\arraystretch}{1.3}
\begin{tabular}{@{}l c c c c c c c c c c c}
\hline
  $R_{in}^b$ & $R_{out}$  &  $R_{out}-R_{in}$ &   $T_d$ & $L_d$ & $q$ & $\chi^2$  &\\
  (R$_{\odot}$) &  (R$_{\odot}$) &  (R$_{\odot}$) & (K) & ($\times 10^{-5}$ L$_{\odot})$& & & \\ 
\hline 
 $0.15^{+0.05}_{-0.03}$  &  $0.23^{+0.05}_{-0.03}$ & $0.08^{+0.03}_{-0.01}$  &  $1280^{+115}_{-94}$  & $7.26^{+0.98}_{-0.83}$ & $0_{-0.6}$ & 14.90 &\\
\hline
\end{tabular}
\end{center}
\end{minipage}
\begin{minipage}{148 mm}
$^a$~Dust particle radius of  0.5~$\mu$m.
$^b$~The smallest possible inner radius is fixed at 0.12 R$_\odot$ in all computations (see text).\\

\end{minipage}
\end{table*}

\section{Results and discussion}
\label{results}
To find the best-fit solution, we varied the disc inner and outer radii between 0.12 and 5.0~R$_\odot$ with a step of 0.01~R$_\odot$ (the case $R_{in} = R_{out}$ was excluded from the computations). The minimum radius is fixed by the size of the central substellar dwarf, which according to the evolutionary models of \citet{Chabrier00} is 0.12~R$_\odot$ for solar metallicity and the age of the system. The index of the disc temperature distribution was also varied between $q = -0.8$ and $q = 0.0$ with a step of 0.10, which mimics the temperature function of a flat protoplanetary disc \citep{ChG97} and a constant temperature, respectively. Negative values of $q$ imply that the inner rim of the disc is hotter than the rest of the disc, which is expected for debris discs. Our fiducial model predicts temperatures of the disc inner rim near the dust sublimation threshold at the regularly accepted values ranging from $\approx 1200$~K for silicate dust particles to $\approx 2000$~K for carbon species \citep{monnier02,Kobayashi09}. Very warm inner discs with temperatures related to the expected dust sublimation temperature for typical dust compositions are known to exist around T\,Tauri stars (e.g. \citealt{muzerolle03b}).  Debris discs with very hot, small dusty particles ($\sim$1500 K, $<$1 $\mu$m) in the inner regions have also been discovered via near-infrared $K$-band interferometry around the Vega star and other A- to K-type stars \citep{absil06,absil13}.

We also varied the total mass of the disc, $M_d$, and the characteristic grain size, $a$. Because there is no confirmed debris material surrounding any known substellar dwarf, we did not have a proper guess for the grain size and the mass of the putative dusty disc of G\,196-3\,B. We varied $M_d$ in the interval $10^{-14}$ M$_{\oplus}$ through 1~M$_{\oplus}$, thus covering 14 orders of magnitude in mass with steps of $(1, 2, ..., 9)\times10^{-14}$~M$_{\oplus}$, $(1, 2, ..., 9)\times10^{-13}$~M$_{\oplus}$, ..., $ (0.1, 0.2, ..., 0.9, 1)$~M$_{\oplus}$. These limits were chosen on the assumption that any putative debris disc should not be less massive than the smallest value accepted for the total mass of Jupiter rings, i.e. $\approx 10^{11}$~kg \citep{Burns04}, and that only about 1~per~cent~of the initial protoplanetary disc mass (expected to be of the order of a few per~cents of the central brown dwarf mass, \citealt{Mohanty13}) remains as a debris disc (1/100 dust-to-gas mass ratio). As a first approximation, we initially adopted a characteristic particle radius of  0.5 $\mu$m for the dust within the disc; similar grain sizes are typically accepted for the interstellar medium. A discussion on the impact of changing this parameter on the debris disc case is provided in Section~\ref{grain radius}.

\subsection{Best-fit solution}
\label{best_fit}
We searched for the minimum value of $\chi^2$, which is defined next and compares the theory with the observations, to find the best solution for our proposed model: 
\begin{equation}\label{formX2}
 \chi^2=\sum^{n}_{j=1}\Big({\frac{f_{\lambda_{j,obs}}-f_{\lambda_{j,mod}}}{\sigma_{j,obs}}}\Big)^2,
\end{equation}
where $f_{\lambda_{j,obs}}$ and $f_{\lambda_{j,mod}}$ stand for observed and predicted flux densities at different observing wavelengths ($\lambda_j$) from 1 up to 24 $\mu$m ($n = 8$). The quantity $\sigma_{j,obs}$ represents the measurement uncertainty for each wavelength. We discarded the shortest wavelengths (below the $J$-band) from the calculation of $\chi^2$ because, as discussed by \citet{ZO10}, the optical energy distribution of G\,196-3\,B is reasonably well reproduced by field L2--L3 dwarfs and the observed flux excesses are noticeable starting at 1.6~$\mu$m.

Table~\ref{table1} provides the derived disc parameters ($R_{in}$ and $R_{out}$) of the best-fit solution (face-on disc) along with the computed minimum $\chi^2$ values. Also provided are the bolometric luminosity, $L_d$, and temperature, $T_d$, of the disc. The bolometric luminosity was obtained by integrating the disc energy distribution employing equation~\ref{form1}, and the bolometric temperature was calculated as follows:
\begin{equation}
 T_d = 1.25\times10^{-11} \frac{\int_{0}^{\infty} \nu~f_{\nu,disc}~{\rm d}\nu}{\int_{0}^{\infty} f_{\nu,disc}~{\rm d}\nu} \;\; \mathrm{(K)}, 
\end{equation}
that represents the temperature of a black body of the same mean frequency \citep{MyersLadd1993}. The integrals were performed over the wavelength range 0.3--100~$\mu$m and using the Simpson's rule. 

The error bars indicated in Table~\ref{table1}  correspond to the 1~$\sigma$ uncertainties obtained from the  $\Delta\chi^2$ confidence statistics by exploring the $\chi_{mod}^2$ surface function\footnote{To obtain adequate $\Delta\chi^2$  boundaries in the multi-dimensional parameter space, we re-scaled each $\chi^2$ value by introducing $\chi_{mod}^2 = DoF \times \chi^2$ / $\chi_{min}^2$, where $DoF$ stands for the number of degrees of freedom of the model and $\chi_{min}^2$ is the global best-fit minimum.} of the fitted parameters in the models. The error in the luminosity of G\,196-3\,B was also considered in the calculations. We caution that the error bars also account for possible variations of the dust grain radius (see further discussion in Section~\ref{grain radius}).

\begin{figure}
\centering
\includegraphics[width=8.5cm]{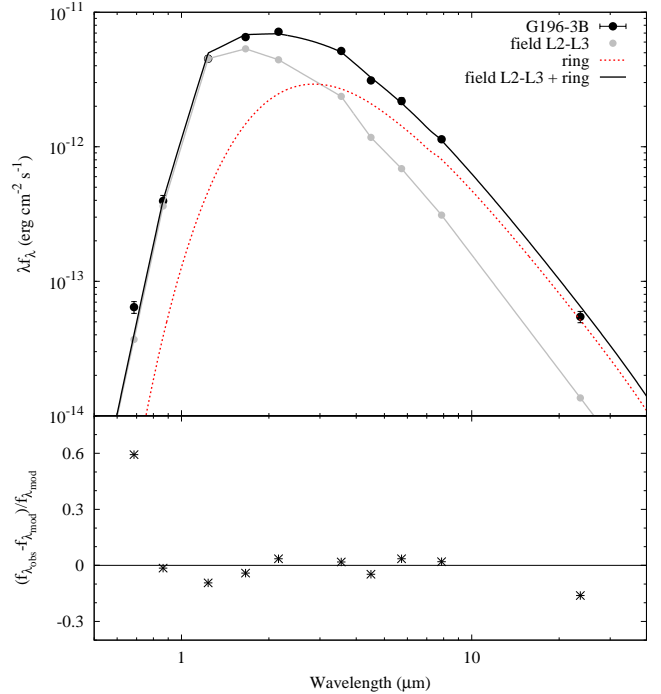} 
\caption{{\sl (Top.)} The observed photometric SED of G\,196-3\,B (black dots, \citealt{ZO10}) is shown in comparison with the best-fit face-on debris disc model (black solid line). The photospheric emission of field L2--L3 dwarfs normalized to the $J$-band (1.25~$\mu$m) of the target is depicted by the gray dots and solid line. The flux emission from the disc parametrised by the best-fit solution is shown with a red dotted line. {\sl (Bottom.)} Normalized residuals as a function of wavelength.}
\label{fig1}  
\end{figure}

The modelled best-fit solution (face-on geometry) is plotted in comparison with the observed SED of G\,196-3\,B in the top panel of Fig.~\ref{fig1}. The bottom panel displays the differences computed as $(f_{\lambda_{obs}}-f_{\lambda_{mod}})/f_{\lambda_{mod}}$ for each wavelength between the predicted and the observed fluxes. The best-fit model suggests the presence of a narrow warm disc, resembling a ``ring", very close to the central substellar source. The ring would have a bolometric temperature of $1280^{+115}_{-94}$~K and would be located in a $0.08^{+0.03}_{-0.01}$~R$_{\odot}$ ($\approx$0.67~$R_{\ast}$) width annulus at the distance of $0.15^{+0.05}_{-0.03}$~R$_{\odot}$ ($\approx$1.25~$R_{\ast}$) from the central dwarf. The computations of the optical depth of the disc material using Equation~\ref{form3} indicate that the resulting warm debris ring mostly comprised of 0.5~$\mu$m particles is optically thick for all of the wavelengths of our study. Hence, the model is not sensitive to the total mass of the disc, and we can only infer the minimum mass required to fit the observations. As listed in Table~\ref{table1}, the disc minimum mass is estimated at $M_d \ge 7.00_{-4.00}\times10^{-10}$~M$_{\oplus}$ (where M$_{\oplus}$ denotes the Earth's mass) for our fiducial model. Equations also provide the disc luminosity ($L_d$, shown in Table~{\ref{table1}) that, when combined with the photosperic luminosity of the central brown dwarf ($L_\ast$), sums up the total $L$ of G\,196-3\,B with a modest deviation of $0.026\times10^{-4}$~L$_\odot$ (see Section~\ref{model}), which is smaller than the quoted luminosity uncertainties.

One striking result of our modelling effort is that the index of the disc temperature power law  distribution (Eq.~\ref{form10}), $q$, is null for the best-fit solution, suggesting that the part of the disc contributing to the observed fluxes up to 24 $\mu$m is practically isothermal across all of its extension. This is an unusual $q$ value for debris and even protoplanetary discs, where the temperature is expected to change radially with the distance to the central heating source (e.g., \citealt{Dullemond01}). Possible interpretations for our result are the following: (i) the detected red emission is dominated by and originates from warm dust particles located at the same distance from the dwarf in the form of, e.g., a dusty spherical envelope or a puffed-up inner rim (wall) with significant scale height. None of these scenarios seems plausible because the optical photometry and spectroscopy of G\,196-3\,B are inconsistent with the central dwarf being heavily extinguished. This indicates that the central dwarf is likely free of optically thick dust material in most of its surface. (ii) Alternatively, because the best-fit solution yields that the optically thick and isothermal materials lie very close to the central object (see below), the upper atmosphere of the brown dwarf may be smudged with dusty spots or dust bands may be present along the surface. And (iii), the $q$ parameter has a large associated uncertainty, which suggests that we cannot confidently constrain the temperature distribution of the disc with the current data. In Table~\ref{table2}, we provide various model computations for different values of $q$. As seen from the $\chi^2$ column, $q$ between 0.0 and $-0.4$ produces quite similar solutions for a characteristic particle size of 0.5 $\mu$m, although only $q = 0.0$ strictly generates the smallest $\chi^2$ (Table~\ref{table1}). Within this range, all the disc parameters change slightly with $q$. The disc gets broader and the inner radius gets closer to the central dwarf with decreasing $q$ because the disc has smaller temperatures at further radii. More negative values of $q$ yield unacceptable results (and are not shown in Table~\ref{table2}) where the combined disc and central dwarf luminosities differ by more than 1 $\sigma$ with respect to the measured total luminosity. The effect of changing  $q$ for fixed disc parameters is discussed in Appendix~\ref{comments} and illustrated in Fig.~\ref{figA1}. }

\begin{table}
\begin{minipage}{84 mm}
\caption{ Best-fit face-on disc parameters as a function of temperature index $q$ (fixed grain size $a$ = 0.5~$\mu$m).}
\label{table2}
\setlength{\tabcolsep}{6.5pt}
\renewcommand{\arraystretch}{1.}
\begin{tabular}{@{}ccccccccc}
\hline
$q$ & $R_{in}$ & $R_{out}$  &   $T_d$ & $M_d$$^{a}$ & $L_d$$^{b}$  & $\chi^2$ & \\
 &  (R$_{\odot}$) &  (R$_{\odot}$)  &  (K) & (M$_{\oplus}$) & (L$_{\odot}$)& & \\ 
\hline 

0       &  0.15  &    0.23 &     1282 &     $\ge$0.7 &     7.26 &         14.90 & \\
$-0.1$  &  0.13  &    0.21 &     1315 &     $\ge$4.0 &     7.39 &         17.14 & \\
$-0.2$  &  0.13  &    0.22 &     1285 &     $\ge$0.6 &     7.36 &         15.47 & \\
$-0.3$  &  0.12  &    0.22 &     1284 &     $\ge$0.5 &     7.42 &         18.30 & \\
$-0.4$  &  0.12  &    0.22 &     1232 &     $\ge$2.0 &     6.78 &         24.05 & \\

\hline
\end{tabular}
\end{minipage}
\begin{minipage}{84 mm}
 $^a$~In units of $10^{-9}$.\\
 $^b$~In units of $10^{-5}$.\\
\end{minipage}
\end{table}

Ring systems or at least debris material, although not directly observed yet, are reported to exist around other substellar objects: The anomalously high brightness of the young planet Fomalhaut~b at optical wavelengths and the peculiar light curve of 1SWASP~J140747.93$-$394542.6 may be explained by the presence of complex ring systems or dust clouds \citep{kalas13,kenyon14,kenworthy15}. Additionaly, the recent discovery of a Venus-mass planet orbiting at a separation of $\approx$~0.34~au from a brown dwarf \citep{Udalski15} suggests that small companions can be formed in discs surrounding substellar objects. The parameters of the modelled debris ring around G\,196-3\,B resemble those of the giant planets in the Solar System (see Section~\ref{planets_rings}) better than stellar debris discs.

\subsection{Effect of changing grain size and disc mass}
\label{grain radius}
Up to this point, we modelled the putative debris disc around G\,196-3\,B using spherical grains of characteristic radius $a$~=~0.5~$\mu$m assuming that the particles within the ring could not be smaller than the interstellar medium grains. However, given the age of the system, there might have existed some grain growth as a result of coagulations and disc evolution. Additionally, in debris discs, grains can also result from the destruction processes of large bodies. Furthermore, the size, growth, and shape of the grains may be a function of the separation to the central object. Here, we explored various debris disc scenarios by adopting spherical grain sizes that range from 0.01 to 100~$\mu$m. As in Section~\ref{model}, we did not include grain dynamics for simplicity. Table~\ref{table3} provides the best-fit inner and outer ring radii for each grain size along with the computed disc mass, bolometric luminosity and temperature, index of disc temperature distribution, and the minimum $\chi^2$. As shown in Table~\ref{table3}, all simulations yielded rather narrow and warm discs (dusty belts) with widths ranging from 0.07 to 0.11~R$_\odot$ and disc luminosities in the interval 6.88--7.34\,$\times$\,10$^{-5}$~L$_\odot$. The $\chi^2$ shows the smallest values for submicron-sized particles suggesting that the putative disc of G\,196-3\,B is most probably made of tiny grains. This result is expected since small grains can be heated to higher temperatures compared to large grains. Increasing the characteristic particle size above 1~$\mu$m produces greater values of $\chi^2$, yet within an acceptable range. Therefore, the current data did not allow us to securely constrain one particular characteristic radius of the disc particles. Also from Table~\ref{table3}, we inferred that increasing the grain size has the effect of bringing the disc closer to the central source so that the disc temperature becomes sufficiently high to reproduce the observed emission fluxes at 1--4~$\mu$m. For grains with $a \ge 2$~$\mu$m, the best-fit solutions give very similar results (except for the minimum mass of the disc). This is caused by the method we employed to determine the radiative equilibrium temperature of the inner rim of the disc (Equation~3 of \citealt{Backman93}, which is independent of the grain size for ``large'' particles, see also Section~\ref{model}). The masses of the disc/ring are different because they do related with the opacity and the optical thickness of the disc material (see Equation~\ref{form3}). The dust opacities we used in all computations including characteristic grain sizes from 0.01 to 100~$\mu$m are presented graphically in Appendix~\ref{comments}.

\begin{table}
\begin{minipage}{84 mm}
\caption{Best-fit face-on ring parameters as a function of characteristic grain radius.}
\label{table3}
\setlength{\tabcolsep}{4.8pt}
\renewcommand{\arraystretch}{1.}
\begin{tabular}{@{}cccccccccccc}
\hline
$a$ & $R_{in}$ & $R_{out}$  &   $T_d$ & $M_d$$^{a}$ & $L_d$$^{b}$ & $q$ &  $\chi^2$ & \\
($\mu$m)&  (R$_{\odot}$) &  (R$_{\odot}$)  &  (K) & (M$_{\oplus}$) & (L$_{\odot}$)& & & \\ 
\hline 
0.01      &  0.14 & 0.22  &  1298  & $\ge$~1.0 & 7.34 & 0  & 15.40 & \\
0.1       &  0.14 & 0.22  &  1298  & $\ge$~1.0 & 7.34 & 0  & 15.38 & \\
0.2       &  0.12 & 0.21  &  1291  & $\ge$~1.0 & 7.32 & $-$0.2  & 15.34 & \\
0.3       &  0.18 & 0.25  &  1302  & $\ge$~0.6 & 7.40 & $-$0.1  & 15.37 & \\
0.4       &  0.15 & 0.23  &  1277  & $\ge$~0.9 & 7.20 & $-$0.2  &  15.43 & \\
0.5$^d$   &  0.15 & 0.23  &  1282  & $\ge$~0.7 & 7.26 & 0  & 14.90 & \\
0.6       &  0.13 & 0.22  &  1273  & $\ge$~0.6 & 7.19 & $-$0.1  & 15.33 & \\
0.8       &  0.12 & 0.21  &  1270  & $\ge$~4.0 & 7.03 & 0  & 17.04 & \\
1.0       &  0.12 & 0.23  &  1249  & $\ge$~0.3 & 6.99 & 0  & 21.03 & \\
2.0       &  0.12 & 0.22  &  1223  & $\ge$~0.7 & 6.86 & 0  & 21.80 & \\
5.0       &  0.12 & 0.22  &  1221  & $\ge$~4.0 & 6.88 & 0  & 21.88 & \\
10.0      &  0.12 & 0.22  &  1221  & $\ge$~7.0 & 6.88 & 0  & 21.88 & \\
100.0     &  0.12 & 0.22  &  1221  & $\ge$~60  & 6.88 & 0  & 21.88 & \\
\hline
\end{tabular}
\end{minipage}
\begin{minipage}{84 mm}
$^a$~In units of $10^{-9}$.\\
$^b$~In units of $10^{-5}$.\\
$^d$~Grain radius used to derive the ring parameters of Table~\ref{table1}.
\end{minipage}
\end{table}

\begin{figure}
\centering
\includegraphics[width=8.7cm]{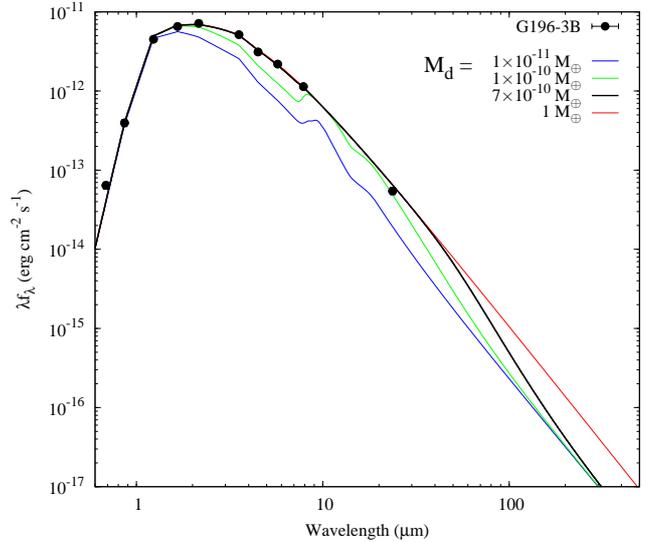} 
\caption{
SEDs computed for different ring masses, from $M_d = 10^{-11}$ through 1~M$_{\oplus}$. Other disc parameters are taken from Table~\ref{table1}. The observed SED of G\,196-3\,B is plotted as black dots.
}
\label{fig2}  
\end{figure}

Additionally, for a given particle size, we inspected the variation of the $\chi^2$ values as a function of the total mass of the disc (between $M_d = 10^{-14}$ to 1.0 M$_{\oplus}$). We found that given the limited amount of observational data of G\,196-3\,B and the optically thick nature of the computed debris belts at these wavelengths, our models are not very sensitive to the total disc mass once a certain critical mass necessary to emulate the 1--24~$\mu$m fluxes is reached. Actually, the best-fit models can determine the minimum mass of the disc required to reasonably reproduce the dwarf's SED; any higher disc mass yields models with similar, although slightly larger, values of $\chi^2$. This is illustrated in Fig.~\ref{fig2}, where SED computations obtained for four different debris ring masses ($M_d$ = $10^{-11}$,  $10^{-10}$, $7\times10^{-10}$, and 1 M$_{\oplus}$) are shown together with G\,196-3\,B's photometry. All other parameters defining the discs were taken from Table~\ref{table1}. It becomes apparent that the debris disc simulations have a marked dependency on small belt masses. For $M_d$ = $10^{-10}$ M$_{\oplus}$, the resulting $\chi^2$ is 405.32, quite far from the best-fit solution that has $\chi^2$ = 14.90 and $M_d$ = $7\times10^{-10}$ M$_{\oplus}$ (Table~\ref{table1}). Greater disc masses, however, do not worsen the fit significantly and yield simulations with slightly larger $\chi^2$ values (15.54 for $M_d$ = 1.0 M$_{\oplus}$). As seen from Fig.~\ref{fig2}, models with massive discs notably differenciate at long wavelengths. To further constrain the total disc mass, data beyond 24 $\mu$m are demanded (see Section~\ref{+disk}). 

We also recall that for simplicity our models employ a fixed characteristic grain radius. For a typical particle size distribution represented by a power law with explonent of 3.5 \citep{Ricci2010a, Ricci2010b} and grains between 0.1 $\mu$m and 1 mm, those with radii below 10 $\mu$m account for roughly 10\%~of the mass of the disc, but are a powerful source of opacity contributing with about 90\%~to the obscuration of the central dwarf ultraviolet and blue light while radiating efficiently at near- and mid-infrared wavelengths. Given the limited amount of data of G\,196-3\,B, our study is ``blind'' to large grains and consequently to the total mass of the disc. To properly determine the total mass of any putative disc surrounding G\,196-3\,B, observations at mid-infrared and radio wavelengths would be required, leading to the exploration of the outer cold debris disc (Section~\ref{+disk}).

The inferred sub-micron characteristic size of the grains within the warm disc of G\,196-3\,B contrasts with the sizeable particles found in other stellar debris disc systems. For the dust belt around the A4-type star Fomalhaut, \citet{min10} found the scattering properties to be consistent with a predominance of $\approx$~100~$\mu$m silicate grains. A similar particle size is obtained by \citet{krivov13} for the cold debris discs surrounding F-, G-, and K-type stars discovered by the space telescope {\sl Herschel}. And particles of about a millimeter are reported to exist in the inner debris disc of the M1V AU Microscopii star, while porous grains sized 0.05--3~$\mu$m populating exterior regions are required to explain the observed blue colour of the scattered light \citep{fitzgerald07}. The great majority of these stellar debris discs have temperatures typically below a few hundred K. 

\subsection{Effect of non-zero inclination}
\label{inclination}
Observed SEDs emerge as the addition of the fluxes of all emitting surfaces/sources forming the system.  Besides the central object and the belt surface seen in face-on configurations (inclination angle $i$ = 0\degr~with respect to the plane of the sky), discs of inclined geometries have two more  emitting superficial regions: The interior and exterior rims assuming truncated discs with cut-off inner and outer radii. To account for these flux sources and the fact that only part of the total surface of the disc is seen when there is an inclination angle $i$, the flux density of equation~\ref{form2} must be multiplied by cos\,$i$, the opacity $\tau$ in equation~\ref{form3} must be multiplied by sec\,$i$, and two new summands are to be included as follows:
\begin{equation}\label{form4}
f_{incl} = f_{\nu,disc}\cos i + f_{\nu,in} + f_{\nu,out},
\end{equation}
where $f_{\nu,in}$ and $f_{\nu,out}$ correspond to the fluxes of the interior and exterior rims. Following a simple approach and under the frame of the black-body approximation, we adopted:
\begin{equation}\label{form5}
f_{\nu,in(out)}=\frac{S_{in(out)}}{d^2} ~B_{\nu}(T_{in(out)}),
\end{equation}
where the subscript $in(out)$ indicates either inner or outer rims, and $S_{in(out)}$ represents the projected emitting area of the inner (outer) rims according to the following expressions \citep{Zakhozhay15}:
\begin{equation}\label{Sin}
S_{in} = 2\,h_{in}\,\sin(i)\sqrt{R_{in}^2-\left(\frac{h_{in}\tan(i)}{2}\right)^2},
\end{equation}
\begin{equation}\label{Sout}
S_{out} = 2\,R_{out}\,h_{out}\,\sin(i),
\end{equation}
where $h_{in}$ and $h_{out}$ stand for the disc half thickness at radii of $R_{in}$ and $R_{out}$, respectively.

To study the impact of an inclination angle on the debris disc case, we fixed the debris ring best-fit parameters according to Table~\ref{table1}. The belt inner radius of 0.15~R$_{\odot}$, G\,196-3\,B's physical size of $R_{\ast}$~=~0.12~R$_{\odot}$, and the maximum vertical half thickness of the disc given by equation~\ref{form33} (see below) allowed us to explore inclination angles from 0 to $\approx$~$30\degr$ without the need to consider the extinction of the central L-type brown dwarf\footnote{An inclination angle of $31\degr$ was determined as the minimum angle after which the disc inner edge starts to shield the central object. This angle depends on the combination of $R_{\ast}$, $R_{in}$, and $h_{in}$ (for more details see \citealt{Zakhozhay15}).}. Our main objective here was to analyse whether an inclined debris disc provides a better fit to the observed data than the face-on geometry. 

In these computations, the two free variables were $i$ and the disc vertical half-thickness $h_d$. The inclination angle was varied between 0\degr~(face-on) and 30$\degr$ with a step of 5\degr. To further simplify the calculations, we adopted a plane geometry, this is a constant vertical thickness throughout the full extension of the disc: $h_{in} = h_{out} = h_d$. This assumption is motivated by the relatively narrow ring (0.08~R$_{\odot}$) found in the previous simulations. The vertical half thickness of the disc was varied between 1~m and approximately $10^7$~m (0.015~R$_{\odot}$, 0.13~$R_{\ast}$) with a step of (1, 2, 3, 4, ..., 9), (1, 2, ..., 9) $\times 10^1$, (1, 2, ..., 9) $\times 10^{2}$, ..., (1, 2, ..., 9, 10) $\times 10^6$~m. The adopted minimum thickness corresponds to the smallest grossness of Saturn's rings \citep{reffet15}, while the maximum explored height is parametrised from the expression given by \citet{KH87}:

\begin{equation}\label{form33}
h_d=0.1\,R_{\ast}\left(\frac{R_{in}}{R_{\ast}}\right)^{9/8},
\end{equation}
where $R_{in} = 0.15$~R$_\odot$ and $R_{\ast} = 0.12$~R$_\odot$ yields a half thickness of 0.015~R$_{\odot}$. Equation~\ref{form33} is commonly used to parametrise protoplanetary flaring disc heights and is consistent with a vertically isothermal gas-rich disc that can pressure-support dust at high disc altitudes. By adopting its value, we were assuming that any debris disc around G\,196-3\,B cannot be thicker than the corresponding protoplanetary disc at a given separation from the central source. For the longest ring scales of our best-fit simulations, this equation fixed a maximum half thickness at about one third to a half of the central brown dwarf's radius.

\begin{figure}
\centering
\includegraphics[width=8.8cm]{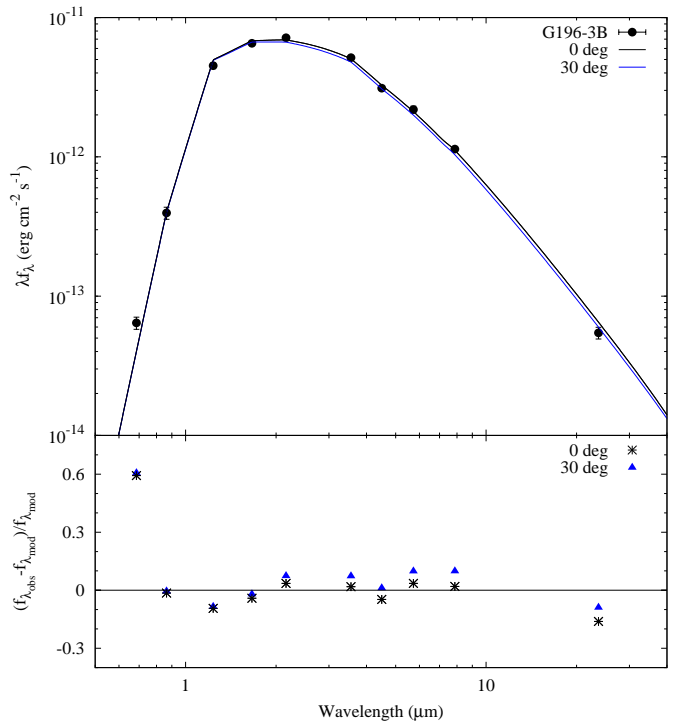} 
\caption{{\sl (Top.)}~The SED of G\,196-3\,B is depicted (black dots) along with the best-fit computations with face-on orientation (black solid line) and with inclination angle  $i$ = 30\degr~(blue solid line). The inner and outer belt radii are 0.15 and 0.23~R$_{\odot}$, the disc half thickness is 0.004~R$_{\odot}$, and the characteristic particle size is 0.5~$\mu$m. {\sl (Bottom.)}~Normalized residuals as a function of wavelength:  $i$ = 0\degr (black asterisks) and 30\degr (blue triangles).}
   \label{fig3}  
\end{figure}

The best-fit solution provided by Eq$.$~\ref{formX2} yielded a debris belt with an inclination angle of $i$ = 5\degr and a vertical half thickness of $h_d = 3\times10^{6}$~m (0.004~R$_{\odot}$ or 0.036~$R_{\ast}$). The minimum $\chi^2$ was found to be 14.89, negligibly smaller than that of the face-on geometry (see Table~\ref{table1}). The top panel of Fig.~\ref{fig3} illustrates the effect of the inclined geometries on the computed SED profiles for $i$ = 0\degr and 30\degr, and the bottom panels shows the residuals between models and observations similarly to Fig.~\ref{fig1}. All other parameters, like grain size, inner and outer disc radii, and disc mass are as indicated in Table~\ref{table1}. Both SEDs are quite alike: the inclined disc model produces smaller fluxes at nearly all wavelengths of study, probably because this disc has a reduced projection of the emitting area in the direction toward the observer. The limited amount of available data and the likely narrow width of the ring prevented a better constraint of the inclination angle of the system.

In Appendix~\ref{comments}, we included graphs to visually illustrate the dependence of the debris disc model on various physical parameters, like the radial temperature distribution power law, the inner and outer radii and the width of the disc.

\subsection{Comparison with dusty model atmospheres}
To compare the best-fit solution found for the observed SED of G\,196-3\,B with theoretical model atmospheres, we retrieved the BT-Settl synthetic spectrum computed for $T_{\ast}$ = 1800~K, surface gravity of log\,$g$ = 4.5~[cm\,s$^{-2}$], and solar metallicity \citep{Allard03,Allard12}, which are close to the accepted physical parameters of the brown dwarf. The BT-Settl models include all of the significant sources of atmospheric gas opacity, the condensation of certain refractory elements in layers where physical conditions favour such condensation, and the opacity due to the new condensate species. According to the theory of model atmospheres (see also \citealt{allard01,marley02,burrows06}), silicate dust grains certainly form in significant quantities in the outer atmospheric layers of L-type objects producing a greenhouse effect and an absorption at blue wavelengths. This induces red colours. After temperature and metallicity, gravity is possibly the most important parameter that defines the colours of L and T dwarfs. As discussed by \citet{burrows06} and \citet[and references therein]{leggett07}, low gravities and small particle sizes in the atmospheres lead to slightly redder infrared indices than high gravities and large grains. For the great majority of field L dwarfs, novel model atmospheres appear to reproduce the observed colours reasonably well. However, there are very red L dwarfs, many of which are relatively young like G\,196-3\,B, which cannot be easily emulated by the theory of atmospheres. 

\begin{figure}
\centering
\includegraphics[width=8.7cm]{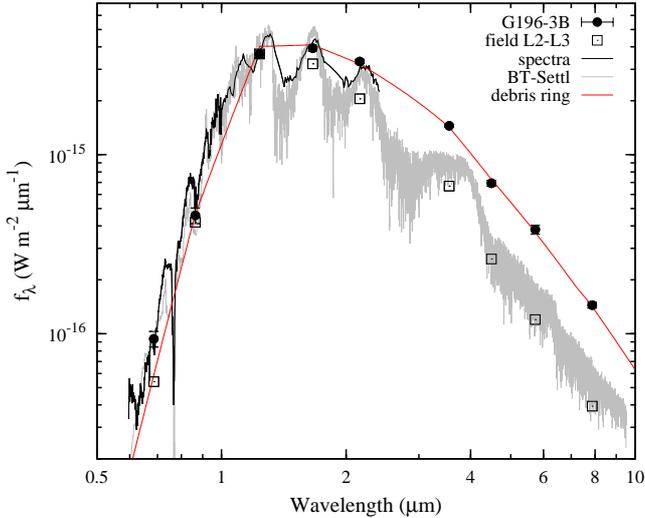}
\caption{ Spectrum and photometric data of the L3 dwarf G\,196-3\,B (black solid line and dots) is compared with the BT-Settl spectrum (gray line) of solar metallicity, log\,$g$ = 4.5~[g\,cm$^{-2}$], and $T_\ast$ = 1800~K \citep{Allard03,Allard12}) normalized to the $J$-band of the observations. The photospheric emission of field L2--L3 dwarfs is shown with open squares. The red line denotes our best-fit model with the characteristics given in Table~\ref{table1}.}
\label{fig4}  
\end{figure}

Fig.~\ref{fig4} shows the combined photometric and spectroscopic data of G\,196-3\,B (as in \citealt{ZO10}). Over-plotted are the spectrum of field L2--L3 objects, the computed fluxes including the contribution of a debris disc with the parameters given in Table~\ref{table1} and the prescription of Section~\ref{model}, and the intermediate-gravity, 1800~K BT-Settl spectrum. The former and the latter data are normalised to the $J$-band of G\,196-3\,B observations. Fig.~\ref{fig4} displays the wavelength interval 0.6--10~$\mu$m, which is the longest wavelength provided by the model atmosphere. In the optical regime ($\lambda < 1$~$\mu$m), the agreement between the BT-Settl model and the observed spectrum is noticeable. However, the discrepancies between the synthetic spectrum and G\,196-3\,B's fluxes get increased for wavelengths $\lambda > 2$~$\mu$m. The debris ring model reproduces the observations to a better level of achievement. Modellers of atmospheric spectra will shortly release improved synthetic spectra (Allard 2015, priv$.$ communication).

\begin{table*}
\begin{center}
\caption{Physical parameters of the rings of Solar System giant planets (excluding halo rings), and best-fit debris ring of G\,196-3\,B.}
\label{table4}
\setlength{\tabcolsep}{4.5pt}
\renewcommand{\arraystretch}{1.}
\begin{tabular}{lccccccccccc}
\hline
Object & $R_{in}^{(1)}$  & $R_{out}^{(1)}$ & $R_{in}^{(1)}$ & $R_{out}^{(1)}$ & $M_d$          & $M_{\rm pl}^{(2)}$  &   $d_{\rm grain}$ & $2h_d$               & $2h_d$ \\
       & (R$_{\odot}$)   & (R$_{\odot}$)   & ($R_{\rm pl}$) &  ($R_{\rm pl}$) & (M$_{\oplus}$) & (M$_{\oplus}$)      &   ($\mu$m)        & (km)               & ($R_{\rm pl}$)    \\
\hline 
G\,196-3\,B & 0.15 &  0.23  &  1.25  &  1.92  & $\ge 7\times10^{-10}$  & 4994.3 &  1   &  $\sim$6,000   &   $\sim$0.072 \\
Jupiter     & 0.13 &  0.32  &   1.30  &  3.15  & $1.67\times(10^{-14}-10^{-9}$) $^{(3)}$ & 317.8  &  $\sim$ 15 $^{(4)}$          & $\sim$80--4,400 $^{(4,5)}$ &  $\sim$0.001--0.062 \\
Saturn      & 0.10 &  0.69  &   1.11  &  8.00  & $(2.5-3.3)\times10^{-6}$ $^{(1)}$       &  95.1  &  $\sim 1 - 10^6$ $^{(1)}$    & $\sim$0.002--17 $^{(8,9)}$   &  $\sim (0.0003-2)\times10^{-4}$\\
Neptune     & 0.06 &  0.09  &   1.69  &  2.54  & $\sim 10^{-13}$ $^{(6)}$                &  17.2  &  $\sim 1 ^{(6)}$             &                            &  \\
Uranus      & 0.06 &  0.07  &   1.49  &  1.95  & $\sim 10^{-9}$ $^{(1)}$                 &  14.5  &  $1\pm0.3$ $^{(7)}$          & $\sim$10--900 $^{(10)}$     &  $\sim 0.0004-0.035$ \\

\hline
\end{tabular}
\end{center}
\medskip
$^{(1)}$ \citet{Vidmachenko2012}, except for G\,196-3\,B; 
$^{(2)}$ \citet{Allen81}, except for G\,196-3\,B;
$^{(3)}$ \citet{Burns04};
$^{(4)}$ \citet{Throop04};
$^{(5)}$ \citet{esposito02}; 
$^{(6)}$ \citet{Miner07};
$^{(7)}$ \citet{Ockert87};
$^{(8)}$ \citet{reffet15};
$^{(9)}$ \citet{scharringhausen13};
$^{(10)}$ \citet{depater13}.
\end{table*}

\subsection{Presence of an outer disc?}
\label{+disk}

The best-fit models obtained in previous Sections account for only one warm (face-on and inclined) ring capable of reproducing G\,196-3\,B's observed fluxes at $\lambda \leq 24$~$\mu$m. However, material at temperatures lower than those of the warm belt mostly emits at longer wavelengths, where there are no available data for the young L-type brown dwarf. G\,196-3\,B might also host cold dust as part of a putative cooler, outer disc. Here, we explored this possibility by taking into account that any cold outer disc has to contribute to wavelengths longer than 24 $\mu$m because, as seen in Fig.~\ref{fig1}, the modelled SED for the best-fit debris disc slightly exceeds the observations at 24~$\mu$m and hence no additional emission at this wavelength could possibly improve the fit.

The existence of structures of two or more debris rings with different temperatures circling stars has been investigated in the literature (e.g. \citealt{Hughes11,ertel11,Donaldson13}). To fit the observational data of A- and G-type stars, these authors introduced two components in their models: a cold debris disc and a ``hot" inner ring. In addition, the giant planets of our Solar System are known to have multiple dusty belts. And \citet{kenworthy15} identified 37 rings around the substellar companion to 1SWASP\,J140747.93$-$394542.6; this particular belt system extends out to a radius of 0.6~au (or 129~R$_\odot$). We simulated a scenario of two rings for G\,196-3\,B consisting of an inner belt with the characteristics given in Table~\ref{table2} for a disc temperature distribution exponent $q=-0.4$, and an outer disc with similar material properties to those described in Section~\ref{model} but located further away from the central L-type dwarf. We adopted $q = -0.4$ because it is physically unlikely that an outer ring keeps a constant temperature, particularly if the outer disc extends to long distances. Furthermore, as discussed in Section~\ref{best_fit}, $q=-0.4$ provides a model with a $\chi^2$ close to that of the best-fit solution and within 1-$\sigma$ the quoted uncertainties. This exponent is also similar to those widely used for debris disks in the literature \citep{Backman93,Hughes11}. 

\begin{figure}
\centering
\includegraphics[width=8.6cm]{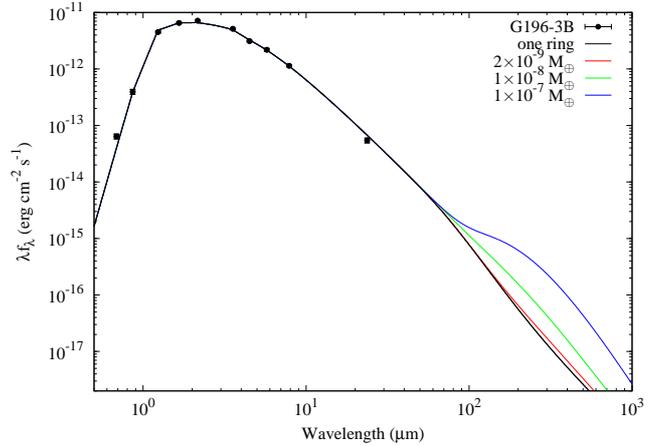} 
\caption{The photometric SED of G\,196-3\,B (dots) is compared with computations considering one inner debris ring (black solid line, belt parameteres as in Table~\ref{table1}) and a combination of an inner ring of different masses and an outer, colder debris disc (red, green, and blue lines). The legend of the figure indicates the masses adopted for the inner belts.}
   \label{fig5}  
\end{figure}

Figure~\ref{fig5} displays examples of computed SEDs with two debris rings. For simplicity, we fixed the outer edge of the cold disc at 100~au, thus compromising the maximum flux from any putative outer disc. We also imposed the surface density law defined in equation~\ref{form9} on the outer disc, where $\varSigma_{in}$ is extracted from expression~\ref{Sigma_r} and from the total mass of the inner ring contained within $R_{in}$ and $R_{out}$ (Table~\ref{table2}, $q=-0.4$). Various two-ring models were computed by adopting different masses of the inner debris belt ($M_d \ge 2\times10^{-9}$ M$_{\oplus}$). From Fig.~\ref{fig5} we concluded that if any outer disc exists, it would have to be located at separations typically greater than 20~au from the central brown dwarf, and would have a likely temperature of 30--40~K. As illustrated in Fig.~\ref{fig5}, the shape of the two-ring SEDs changes significantly at wavelengths $\ge$80~$\mu$m with the adopted mass of the inner belt. If there is a vast debris disc around G\,196-3\,B, we were able to study only the innermost region with the current observations. Additional data at wavelengths $>$24~$\mu$m are highly required to firmly investigate the presence of an outer and colder disc.

\subsection{Comparison with giant planets rings}
\label{planets_rings}
In previous Sections, we briefly mentioned the similarities and differences between the best-fit dusty ring of G\,196-3\,B and the debris discs found around stars. Here, we compared the physical properties of the G\,196-3\,B system according to our basic computations with those of the dusty belts circling the giant planets of the Solar System. This comparison is worthy since the most likely mass and radius of the L dwarf are about 15 and 1.2 times those of Jupiter, respectively. Furthermore, according to evolutionary substellar-mass models, G\,196-3\,B will likely have a temperature of $\approx$~300--400~K and the radius of Jupiter at the age of the Solar System, which is only a factor of two warmer than the current temperature of the giant planet. In mass, size, and evolution, G\,196-3\,B is likely more related to Jupiter than to solar-type stars.

As for the giant planets of the Solar System, Table~\ref{table4} provides the inner and outer radii ($R_{in}$, $R_{out}$) and the full vertical thickness ($2h_d$) of their ``complete" systems of rings (excluding halo rings), both in solar units or km and in units of the corresponding planet radius ($R_{\rm pl}$), the disc and central object masses ($M_d$ and $M_{\rm pl}$ in Earth units), and the typical diameter of the grains ($d_{\rm grain}$). We also facilitate the references from which we extracted the relevant information. Our tentative findings for G\,196-3\,B are included in Table~\ref{table4} for a proper analogy. The temperatures of the rings are not shown in Table~\ref{table4} because they are probably  very different considering the distinct effective temperatures of the central objects, and only the temperature of the rings of Saturn is available in the literature to the best of our knowledge. Interestingly, the temperature of Saturn's inner ring is $\approx$~80~K,
  which is rather close to the accepted temperature of the planet (similarly to the results of our simulations for G\,196-3\,B). At equinox, the temperature of Saturn's outer rings decreases down to 40~K with the radial distance from the planet \citep{Spilker13}. According to Table~\ref{table4}, the ring to central object mass ratios of G\,196-3\,B, Jupiter, and Neptune are alike within the large uncertainties (a few orders of magnitude) coming from our computations and the poorly determined mass of the Solar System rings. The width (extension) of the rings is narrower in the case of G\,196-3\,B probably because the limited amount of data prevented the analysis of the presence of a colder and outer disc. However, relative to the size of the central bodies, the location of G\,196-3\,B's warm belt resembles those of the giant planets' rings. The characteristic grain size of the dusty particles is also similar for all objects listed in the Table~\ref{table4}. G\,196-3\,B may host a thicker ring than Solar System planets, with the warmer bolometric temperature of the disc in clear contrast with the cold belts of Solar System bodies. Additionally, because of the very distinct temperatures,  we expect quite different compositions (Neptune and Jupiter rings are mostly made of ice).

\section{Conclusions and final remarks}
\label{conclusions}
We investigated the prospect of the presence of a debris disc surrounding G\,196-3\,B, one $\sim 15$~$M_{\rm Jup}$, L3 young brown dwarf with near- and mid-infrared colour excesses, by fitting its observed spectral energy distribution from the visible through 24~$\mu$m. The computations showed that the debris disc model is capable of delivering physically possible solutions. The best-fit solution yields a warm ($\approx$~1280~K), quite narrow, optically thick debris ring (width of $\approx$~0.07--0.11~R$_\odot$) located very close to the central brown dwarf (separation of $\approx$~0.12--0.20~R$_\odot$). The model also suggests a debris disc mass $\ge 7\times 10^{-10}$~M$_{\oplus}$ containing sub-micron/micron characteristic dusty particles with temperatures near the sublimation threshold of silicates, but below the sublimation temperature of carbon particles. Considering the derived location and size of the belt and the disc-to-brown dwarf mass ratio, the dusty ring around G\,196-3\,B resembles the rings of Neptune and Jupiter (except for the high temperature and likely thick vertical height of $\approx 6 \times 10^3$~km). 

With the data currently available, some parameters, like the surface density profile of the disc and the inclination angle of the system, cannot be properly constrained, and only warm and inner discs have been modelled in detail; to study the presence of outer, colder discs, data at wavelengths longer than 24~$\mu$m are required. The debris disc scenario provides a better reproduction of G\,196-3\,B's SED than the solar-metallicity, low-gravity BT-Settl model atmosphere corresponding to the effective temperature of 1800 K (typical of L3 dwarfs). The existence of debris discs surrounding intermediate-age, red L-type dwarfs may account for the colour flux excesses at near- and mid-infrared wavelengths, and may represent an alternative scenario to that of low-gravity, enshrouded atmospheres.

The origin of this putative debris belt around G\,196-3\,B is unknown. It might result from the direct leftovers of the original disc made of gas and dust. At the time when the gas-to-dust ratio was of the order of 0.1--10 in the disc, the ``small" amount of gas might have organised the dusty particles into a narrow ring \citep{LyraKuchner2013}. Planets or proto-planets formed from the disc material sweep their orbits and may accumulate debris at certain locations within the planetary system (similarly to the Kuiper Belt in the Solar System), or produce a cut-off of the outer edge of the rings (e.g. \citealt{Thebault2012,Ertel12}). Alternatively, the hot dust component may arise in an asteroid belt undergoing collisional destruction or in massive collisions in ongoing terrestrial planet formation. In the latter picture, the high luminosities and temperatures of the debris material resulting from terrestrial planet formation may survive for around a few to several ten~Myr after the impacts \citep{jackson12}, with considerable levels of emission at $\sim$24~$\mu$m that might be readily seen above the intrinsic, faint photospheric flux of low-luminous and cool central objects. 

\section*{Acknowledgements}

Authors thank the anonymous referee for useful suggestions that improved the manuscript.
Authors thank Dr. T. Trifonov for the useful comments on the paper. This work is partly financed by the Spanish Ministry of Economy and Competitivity through the project AYA2014-54348-C3-2-R.



\appendix
\section{}
\label{comments}
\begin{figure*}
\centering
 \includegraphics[width=5.in]{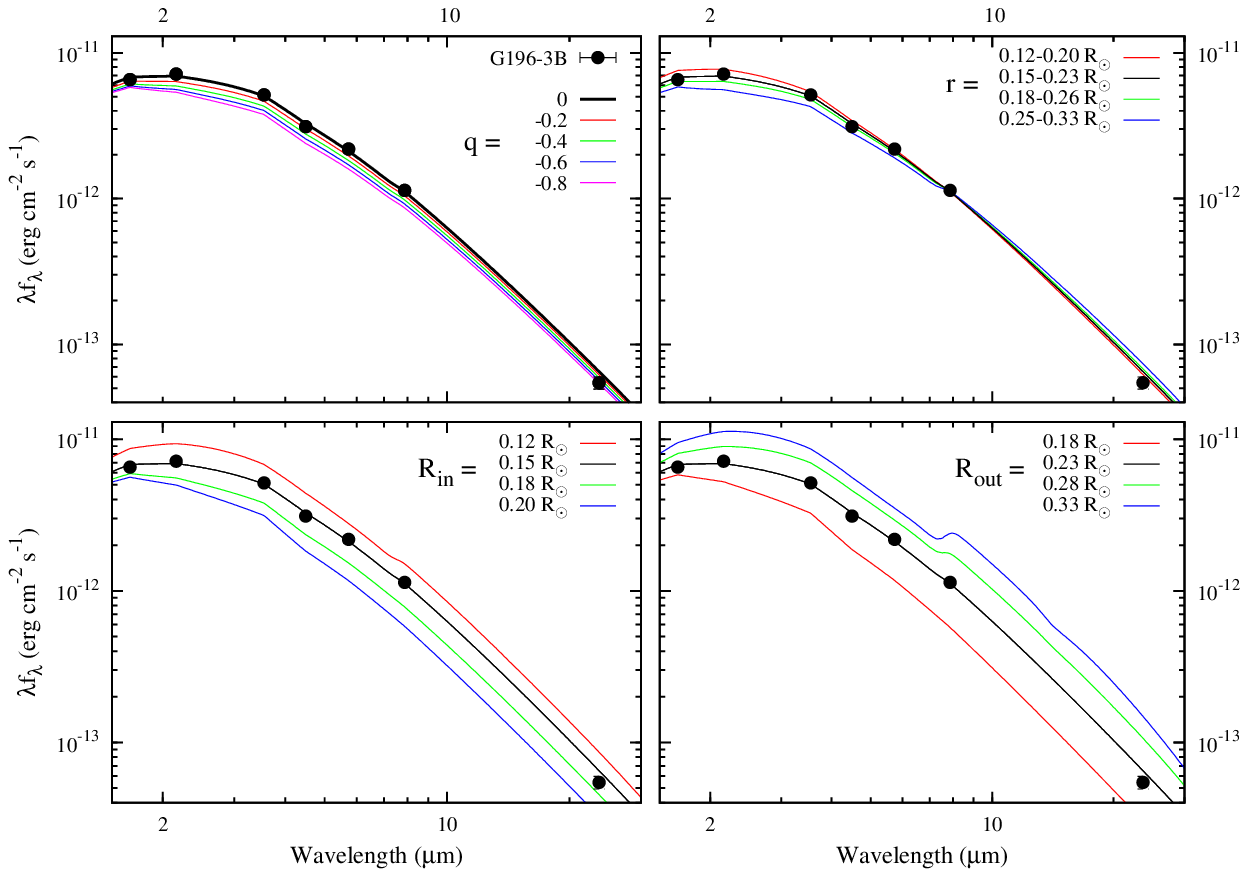} 
 \caption{The spectral energy distribution of G\,196-3\,B (black dots, \citealt{ZO10}) is shown in comparison with theoretical computations for systems with differing disc parameters. The various panels show the dependency of the modelled SEDs on the radial temperature profile index ({\sl top left}), on the inner and outer radii for a fixed disc width ({\sl top right}), on the inner radius for a fixed outer radius ({\sl bottom left}), and on the outer radius for a fixed inner radius ({\sl bottom right}). For each set of models, the simulation providing the smallest $\chi^2$ is plotted with the black solid line.}
   \label{figA1}  
\end{figure*}

In the following, we briefly commented on the qualitative effects of changing some disc parameters within reasonable intervals for disc formalism described in Section~\ref{model}. In Fig.~\ref{figA1} the observed photometric spectral energy distribution (SED) of G\,196-3\,B is compared with various computations produced by varying different parameters while keeping others constant so that the dependency of the simulations on the free variables can be traced graphically. For each set of models, the best-fit solution defined by the smallest associated $\chi^2$ value (Section~\ref{best_fit}) is depicted with a black solid curve. To ease the computations, we imposed a minimum disc inner radius of $R_{\ast}$.

Fig.~\ref{figA1} illustrates the disc simulations according to Section~\ref{model}. The top left panel shows the effect of changing the temperature spectral index $q$ (see equation~\ref{form10}). The remaining three panels depict the changing computed SEDs for varying disc inner radius (and constant outer radius, bottom left panel), varying disc outer radius (and constant inner radius, bottom right panel), and varying disc inner and outer radii (and constant disc width, top right panel). It can be seen that the different $q$ values (from 0 to $-0.8$) have a noticeable effect on the SED profiles. The most significatnt changes occur at $\lambda = 1 - 10$ $\mu$m, where the maximum of the emission from the ring is expected. Increasing the inner radius (including the simulations with a constant disc extension) produces smaller fluxes at red wavelengths ($\lambda = 1 - 24$~$\mu$m), which is explained by a decreasing temperature of the disc. Augmenting the outer radius induces higher  infrared fluxes because of the larger disc emitting area.

\par
\begin{figure}
\centering
  \includegraphics[width=8.7cm]{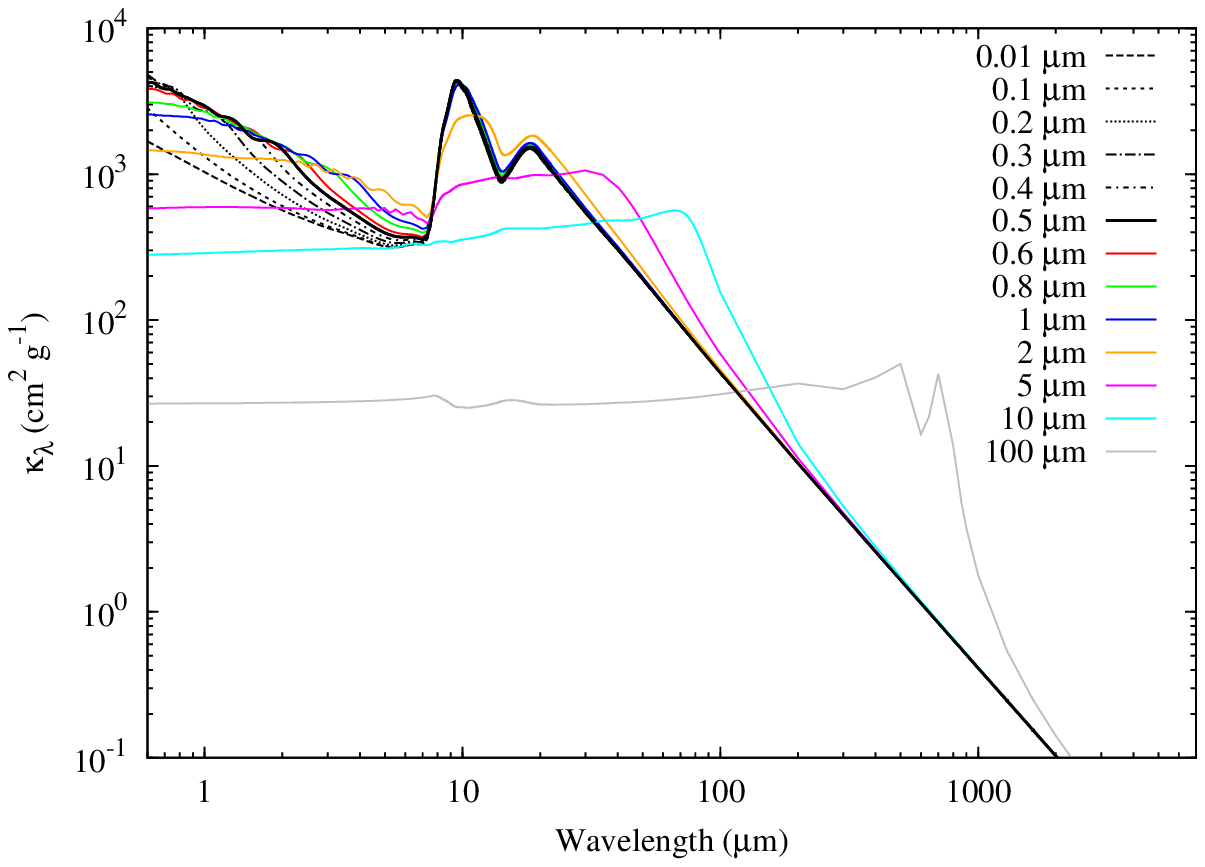} 
 \caption{The opacities for the rings that consist of different characteristic grain radii.}
   \label{figA2}  
\end{figure}

 Fig.~\ref{figA2} presents the dust opacities, computed with Mie theory, for the grain sizes considered in Section~\ref{grain radius}.

\end{document}